\documentclass{article}
\usepackage{graphicx, color, epsfig}
\usepackage{amsfonts}
\usepackage{amssymb}
\usepackage{amsthm}

\newtheorem{definition}{Definition}
\newtheorem{example}{Example}

\newtheorem{corollary}{Corollary}
\newtheorem{lemma}{Lemma}
\newtheorem{theorem}{Theorem}

\newtheorem{property}{Property}
\usepackage{url}

\newcommand{\query}{\mathrm{query}}
\newcommand{\PQRN}{\mathrm{PQRN}}
\newcommand{\add}{\mathrm{add}}
\newcommand{\del}{\mathrm{del}}

\newcommand{\ADD}{\mathrm{ADD}}
\newcommand{\DEL}{\mathrm{DEL}}
\newcommand{\SOP}{\mathrm{SOP}}

\begin{document}

\pagestyle{headings}

\title{Instance-Independent View Serializability for Semistructured Databases}
\author{
  Stijn Dekeyser \\ University of Southern Queensland \\ {\small {\textbf{dekeyser@usq.edu.au}}} \and
  Jan Hidders \\ University of Antwerp \\ {\small {\textbf{jan.hidders@ua.ac.be}}} \and \\
  Jan Paredaens \\ University of Antwerp \\ {\small {\textbf{jan.paredaens@ua.ac.be}}} \and \\
  Roel Vercammen\thanks{Roel Vercammen is supported by IWT -- Institute for the Encouragement of Innovation by Science and Technology
   Flanders, grant number 31581.} \\ University of Antwerp \\ {\small {\textbf{roel.vercammen@ua.ac.be}}}
}

\maketitle

\begin{abstract}
Semistructured databases require tailor-made concurrency control
mechanisms since traditional solutions for the relational model
have been shown to be inadequate.  Such mechanisms need to take
full advantage of the hierarchical structure of semistructured
data, for instance allowing concurrent updates of subtrees of, or
even individual elements in, XML documents. We present an approach
for concurrency control which is document-independent in the sense
that two schedules of semistructured transactions are considered
equivalent if they are equivalent on all possible documents. 
We prove that it is decidable in polynomial time whether two
given schedules in this framework are equivalent. This also solves
the view serializability for semistructured schedules polynomially
in the size of the schedule and exponentially in the number of
transactions.
\end{abstract}

\section{Introduction}

In previous work~\cite{DH02,DH03a,DH03} we have shown that
traditional concurrency control~\cite{WV02} mechanisms for the
relational model~\cite{BHG87,GPT76,P86,SK80} are inadequate to
capture the complicated update behavior that is possible for
semistructured databases.  Indeed, when XML documents are stored
in relational databases, their hierarchical structure becomes
invisible to the locking strategy used by the database management
system.

In general two actions, on two different nodes of a document
tree, that are completely `independent' from each other, cannot
cause a conflict, even if they are updates. Changing the spelling
of the name of one of the authors of a book and adding a chapter
to the book cannot cause a conflict for instance. 
Most classical concurrency control mechanisms, when applied in a
naive way to semistructured data, will not allow such concurrent
updates. This consideration is the main reason why the classical
approaches seem to be inadequate as a concurrency control mechanism
for semistructured data.

Most of the work on concurrency control for XML and
semistructured data is based on the observation that the data is
usually accessed by means of XPath expressions. Therefore it is
suggested in \cite{DH02} to use a simplified form of XPath
expressions as locks on the document such that precisely all
operations that change the result of the expression are no longer
allowed. Two alternatives for conflict-checking are proposed, one
where path locks are propagated down the XML tree and one where
updates are propagated up the tree, which both have their specific
benefits. This approach is extended in \cite{DH03} where a
commit-scheduler is defined and it is proved that the schedules it
generates are serializable. Finally in \cite{DHP03a} an
alternative conflict-scheduler is introduced that allows more
schedules than the previously introduced commit-scheduler.

A similar approach is taken in \cite{CK03} where conflicts with
path locks are detected by accumulating updates in the XML tree
and intelligently recomputing the results of the path expressions.
As a result they can allow more complex path expressions, but
conflict checking becomes more expensive. Another related approach
is presented in \cite{JCW02} where locks are derived from the path
expressions and a protocol for these locks is introduced that
guarantees serializability. 

Several locking protocols that are not based on path expressions
but on DOM operations are introduced in \cite{HKM03,HKM04}. Here, there
are locks that lock the whole document, locks that lock all the
children of a certain node and locks that lock individual nodes
or pointers between them. An interesting new aspect is here the
possibility to use the DTD for conflict reduction and thus
allowing more parallelism. Although these locking protocols seem
very suitable in the case of DOM operations, it is not clear
whether they will also perform well if most of the access is done
by path expressions.  A similar approach, but extended with the
aspect of multi-granularity locking, is presented in \cite{Haustein04a,Haustein04b}.
This approach seems more suitable for hierarchical data like
semistructured data and XML. However, such mechanisms will often
allow less concurrency than a path based locking protocol would.

A potential
problem with many of the previously mentioned protocols is that
locks are associated with document nodes and so for large
documents we may have large numbers of locks. A possible solution
for this is presented in \cite{GBS02} where the locks are
associated with the nodes in a DataGuide, which is usually much
smaller than the document. However, this protocol does not
guarantee serializability and allows phantoms.

For all the approaches above it holds that the concurrency control
mechanisms are somehow dependent upon the document. In most cases
this means that if the document gets very large then the overhead
may also become very large. This paper investigates the
possibilities of a document-independent concurrency control mechanism.
It extends the preliminary results on this subject that were
presented in \cite{DHP03b}.

The total behavior of the processes that we consider in this paper
is straightforward: each cooperating process produces a
transaction of atomic actions that are queries or updates on the
actual document. The transactions are interleaved by the scheduler
and the resulting schedule has to be equivalent with a serial
schedule. Two schedules on the same set of transactions are called
equivalent iff {\bf for each possible input document} they
represent the same transformation and each query gives the same
result in both schedules. This is a special definition of
view equivalency, which we will use to decide view
serializability \cite{bernstein82} for a schedule.

Note that we consider view serializability, as opposed to conflict
serializability. As we will show later on, conflict serializability,
which might be more interesting from a computational point of view,
will allow less schedules to be serialized and hence can be too
restrictive.

The updates that we consider are very primitive: the addition of
an edge of the document tree and the deletion of an edge.
Semantically the addition is only defined if the added edge does
not already exist in the document tree. Analogously the deletion
is only defined if the deleted edge exists. A more general
semantics, that does not include this constraint, can be easily
simulated by adding first some queries.

There are some schedules for which the result is undefined for all
document trees (e.g., a schedule consisting of two consecutive
deletions of the same edge). These schedules are meaningless and
are called inconsistent.  
Hence a schedule is consistent if there exists at least one
document tree on which its application is defined. We prove that
the consistenty of schedules is polynomially decidable. 

In order to tackle the equivalence of schedules and transactions
we first consider schedules without queries, and as such we have
only to focus on the transformational behavior of the schedules.
We will see that, contrary to the relational model,  the swapping
of the actions cannot help us in detecting the equivalence of two
schedules. We prove that the equivalence of queryless schedules is
also polynomially decidable, and that view serializability is
exponentially decidable in the number of transactions and polynomially
in the number of operations.
Finally we generalize the results above for general schedules over
the same set of transactions.

The paper contains a number of theoretical results on which the
algorithms are based. The algorithms are a straightforward
consequence of the given proofs or sketches. The complete proofs
are given in~\cite{HPV04}.

The paper is structured as follows: Section 2 defines the
data model, the operations and the semistructured schedules.
Section 3 studies the consistency of schedules without queries. In
Section 4 we study the equivalence and the view serializability problem
for these queryless schedules. In Section 5 we generalize these
results for consistent schedules.

\section{Data Model and Operations}\label{dmdml}

The data model we use is derived from the classical data model for
semistructured data \cite{ABS99}. We consider directed, unordered
trees in which the edges are labelled.

Consider a fixed universal set of nodes $\mathcal{N}$ and a
fixed universal set of edge labels $\mathcal{L}$ not containing
the symbol $/$.

\begin{definition}
  A {\em graph} is a tuple $(N, E)$ with $N \subseteq \mathcal{N}$
  and $E \subseteq N \times \mathcal{L} \times N$.
  A {\em document tree} ({\sc dt}) $T$ is a tuple $(N, E, r)$
  such that $(N, E)$ is a graph that represents a tree
  with root $r$. The edges are directed from the parent to the
  child.
\end{definition}

\begin{figure}[ht]
\begin{center}
\includegraphics{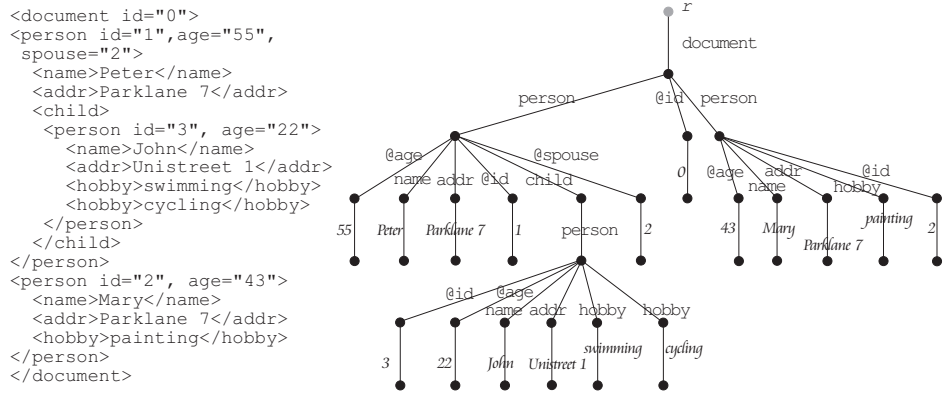}
\end{center}
\caption{A fragment of an XML document and its {\sc dt} representation.} \label{xmlboom}
\end{figure}

\begin{example}
Figure~\ref{xmlboom} shows a fragment of an XML document and its
{\sc dt} representation.
\end{example}

This data model closely mimics the XML data model as illustrated
in the next example. We remark, however, the following differences:
\begin{itemize}
 \item {\bf order:} Siblings are not ordered.  This is
   not crucial, as an ordering can be simulated by using a skewed
   binary {\sc dt}.
 \item {\bf attributes:}  Attributes, like elements, are represented by
   edges labeled by the name of the attributes (started with a @).
   The difference is that in this data model an element may contain several
   attributes of the same name.
 \item {\bf labels:} Labels
   represent not only tag names and attribute names, but also values
   and text. 
 \item {\bf text:}
Unlike in XML, it is possible for several text edges to be
   adjacent to each other.
\end{itemize}

  A {\em label path} is a string of the form $l_1 / \ldots / l_m$
  with $m \geq 0$ and every $l_i$ an edge label in $\mathcal{L}$.
  Given a path $p = ( (n_1, l_1, n_2), \ldots, (n_m, l_m,
  n_{m+1}) )$ in a graph $G$, the {\em label path of $p$},
  denoted $\bar\lambda_T(p)$ (or $\bar\lambda(p)$ when $T$ is
  subsumed) is the string $l_1 / \ldots / l_{m}$.

Processes working on document trees do so in the context of a
general programming language that includes an interface to a
document server which manages transactions on documents.  The
process generates a list of operations that will access the
document. In general there are three types of operations: the
query, the addition and the deletion. The input to a query
operation will be a node and a simple type of path expression,
while the result of the invocation of a query operation will be a
set of nodes. The programming language includes the concepts of
sets, and has constructs to iterate over their entire contents.
The input to an addition or a deletion will be an edge. The
result of an addition or a deletion will be a simple
transformation of the original tree into a new tree. If the result
would not be a tree anymore it is not defined.

We now define the path expressions and the query operations,
subsuming a given {\sc dt} $T$.

The syntax of path expressions\footnote{Remark that path
expressions form a subset of XPath expressions.} is given by
$\mathcal{P}$:
\begin{eqnarray*}
 \mathcal{P} & ::= & pe_\epsilon\ |\ \mathcal{P}^+ \\
 \mathcal{P}^+ & ::= & \mathcal{F}\ |\ \mathcal{P}^+{\tt /}\mathcal{F}\
                   |\ \mathcal{P}^+{\tt //}\mathcal{F} \\
 \mathcal{F} & ::= & *\ |\ \mathcal{L}
\end{eqnarray*}

The set ${\rm \bf L}(pe)$ of label paths represented by a path
expression $pe$ is defined as follows:
\begin{eqnarray*}
   {\rm \bf L}(pe_\epsilon)      & = & \{\epsilon\} \\
   {\rm \bf L}(*)               & = & \cal{L}  \\
   {\rm \bf L}(l)               & = & \{l\}    \\
   {\rm \bf L}(pe {\tt /} f) & = & {\rm \bf L}(pe) \cdot \{{\tt /}\} \cdot {\rm \bf L}(f)    \\
   {\rm \bf L}(pe {\tt //} f)  & = & {\rm \bf L}(pe) \cdot \{{\tt /}\} \cdot({\cal L} \cdot \{{\tt /}\})^* \cdot {\rm \bf L}(f)
\end{eqnarray*}
Let $n$ be an arbitrary node of $T$ and $pe$ a path expression. We
now define the three kinds of operations: the query, the addition
and the deletion.
\begin{definition}
The query operation $\query(n,pe)$ returns a set of nodes, and is
defined as follows:
  \begin{itemize}
    \item $\query(n,pe)$ with $n \in \mathcal{N}$ and $pe \in \mathcal{P}$.
      The result of a query on a {\sc dt} $T$ is defined as
      $\query(n,pe)[T]  =   \{n' \in N\ |\ \exists p \mathrm{\ a\ path \ in\ }T \mathrm{\ from\ } n \mathrm{\ to\ } n' \mathrm{\ with\ }
      \bar\lambda(p) \in {\rm \bf L}(pe)\}$.
  \end{itemize}

  The update operations $\add(n, l, n')$ and $\del(n,l,n')$ return no value but transform a {\sc dt}
  $T = (N, E, r)$ into a new {\sc dt} $T' = (N', E', r)$:
  \begin{itemize}
   \item $\add(n, l, n')$ with $n, n' \in \mathcal{N}$ and $l \in
     \mathcal{L}$. The resulting $T'=\add(n, l, n')[T]$ is defined by $E' = E \cup
     \{ (n,l,n') \}$ and $N' = N \cup \{ n' \}$. If the resulting $T'$ is
     not a document tree anymore or $(n, l, n')$ was already in the
     document tree then the operation is undefined.
   \item $\del(n,l,n')$ with $n, n' \in \mathcal{N}$ and $l \in
     \mathcal{L}$. The resulting $T'=\del(n,l,n')[T]$ is defined by $E' = E -  \{
     (n,l,n') \}$ and $N' = N - \{ n' \}$.  If the resulting $T'$ is
     not a document tree anymore or $(n,l,n')$ was not in the
     document tree then the operation is undefined.
  \end{itemize}
\end{definition}

Note that the operations explicitly contain the nodes upon which
they work. As we will explain in Section~4 this is justified
by the fact that the scheduler decides at run time whether an
operation is accepted or not.

We now give some straightforward definitions of schedules and
their semantics.

\begin{definition}
An {\em action} is a pair $(o,t)$, where $o$ is one of the three
operations $\query(n,pe)$, $\add(n,l,n')$ and $\del(n,l,n')$  and
$t$ is a transaction identifier. A {\em transaction} is a sequence
of actions with the same transaction identifier. A {\em schedule}
over a set of transactions is an interleaving of these
transactions. The size $n_S$ of a schedule $S$ is the length of its
straightforward encoding on a Turing tape\footnote{We assume that
nodes can be encoded in $O(1)$-space}.

We can {\em apply} a schedule $S$ on a {\sc dt} $T$. The result of
such an application is
\begin{itemize}
 \item for each query in $S$, the result of this query.
 \item the {\sc dt} that results from the sequential application of the
   actions of $S$; this {\sc dt} is denoted by $S[T]$
\end{itemize}
If some of these actions are undefined the application is
undefined. Two schedules are equivalent iff they are defined on the
same non-empty set of {\sc dt}s and on each of these {\sc dt}s
both schedules have the same result. The definition of serial and
serializable schedules is straightforward.
\end{definition}

Since a transaction is a special case of a schedule all the
definitions on schedules also apply on transactions.

Note that the equivalence of schedules and transactions is a
document-independent definition. Let \\
$T_1=(\{n_1,n_2\},\{(n_1,l_2,n_2)\},n_1)$, \\
$T_2=(\{n_1,n_2\},\{(n_1,l_1,n_2)\},n_1)$, \\
$T_3=(\{n_1\},\emptyset,n_1)$
be three {\sc dt}s and let \\
$S_1= (\add(n_2,l_2,n_3),t_1), (\query(n_1,l_1/l_2),t_2)$, \\
$S_2=(\query(n_1,l_1/l_2),t_2), (\add(n_2,l_2,n_3),t_1)$
be two schedules. \\
$S_1$ and $S_2$ are equivalent on $T_1$, they are not equivalent
on $T_2$ and their application is undefined on $T_3$. \\
Let $S_3$ be the empty schedule and \\
$S_4= (\add(n_1,l_1,n_2),t_1), (\del(n_1,l_1,n_2),t_2)$. \\
$S_3$ and
$S_4$ are not equivalent although they are equivalent on many {\sc
dt}s.

We will later on use the definition of equivalence to define
serializability. In this paper we study view serializability,
which is less restrictive than conflict serializability. We
illustrate this claim by introducing informally a scheduling
mechanism for generating conflict serializable schedules.
A possible approach for this is to have a locking mechanism
where operations can get locks, and in which a new operation
of a certain process will only be allowed if it does not require
locks that conflict with locks
required by earlier operations.
Because operations with non-conflicting locks can be
commuted, any schedule that is allowed by such a scheduler can be
serialized. The following example shows, however, that the reverse
does not hold:
  Indeed, the next schedule
\begin{tabbing}
    $S =$ \= $(\add(r,l_1,n_1),t_1)$, $(\del(r,l_1,n_1),t_2)$,\\
    \> $(\add(r,l_2,n_2),t_2)$, $(\del(r,l_2,n_2),t_2)$, \\
    \> $(\add(r,l_2,n_2),t_1)$, $(\del(r,l_2,n_2),t_1)$.
  \end{tabbing}

  is consistent since it is defined on
  $T=(\{r\},\emptyset,r)$.
  Furthermore it is serializable, and the equivalent serial
  schedules are
 \begin{tabbing}
    $S_1 =$ \= $(\add(r,l_1,n_1),t_1)$, $(\add(r,l_2,n_2),t_1)$, \\
    \> $(\del(r,l_2,n_2),t_1)$, $(\del(r,l_1,n_1),t_2)$, \\
    \> $(\add(r,l_2,n_2),t_2)$, $(\del(r,l_2,n_2),t_2)$ \\
    $S_2 =$ \> $(\del(r,l_1,n_1),t_2)$,$(\add(r,l_2,n_2),t_2)$, \\
    \> $(\del(r,l_2,n_2),t_2),$ $(\add(r,l_1,n_1),t_1)$, \\
    \> $(\add(r,l_2,n_2),t_1)$, $(\del(r,l_2,n_2),t_1)$.
  \end{tabbing}
but we cannot go from $S$ to $S_1$ nor to $S_2$ only by swapping
with consistent intermediate schedules. This illustrated that an
approach based on conflict serializability can be too strict.

\section{Consistency of Queryless Schedules}
A schedule is called {\em queryless} (QL) iff it contains no queries.
Because of the way that operations can fail it is possible
that the application of a certain transaction is not defined for
any document tree. We are not interested in such transactions. We
call a transaction $t$ {\em consistent} iff there is at least one
{\sc dt} $T$ with $t[T]$ defined.
\begin{example} \label{ex:corrnotcorr} The next transaction is consistent: 

$(\add(r,l_1,n_1),t_1)$,
$(\del(r,l_1,n_1),t_1)$, 
$(\add(r,l_2,n_2),t_1)$, 

$(\del(r,l_2,n_2),t_1)$, 
$(\add(r,l_2,n_2),t_1)$,
$(\del(r,l_2,n_2),t_1)$. \\
Note, however, that there are {\sc dt}s on which this transaction is undefined. 
For example, if $T$ contains an edge $(r, l_3, n_1)$, then $t_1[T]$ is undefined, since
the application of the first action of $t_1$ is undefined. \\
The next transaction is inconsistent: 

$(\add(n_1,l_1,n),t_1)$, $(\add(n_2,l_2,n),t_1)$.
\end{example}

We call a schedule $S$ {\em consistent} iff there is at least one
{\sc dt} $T$ with $S[T]$ defined. Remark that there are consistent
schedules that cannot be serializable because they contain an
inconsistent transaction. For instance, the consistent schedule
 $S =
(\add(r,l_1,n_1),t_1),$ $(\del(r,l_1,n_1),t_2),$
$(\add(r,l_1,n_1),t_1)$
 is  defined on $T=(\{r\},
\emptyset, r)$, and hence is not serializable, because every
equivalent serial QL schedule would be undefined (since the
transaction $t_1$ is not consistent).
Transaction $t_1$ has the property that all QL schedules over
a set of transactions that contain $t_1$ are non-serializable.

Note that the definition of consistent QL schedule is
document-independent. It is clear that we are only interested in
consistent transactions and schedules. Remark also that if two QL
schedules are equivalent then they are both consistent. This
equivalence relation is defined on the set of consistent QL
schedules.

We will characterize the consistent QL schedules and prove that
this property is decidable. For this purpose we will first attempt
to characterize for which document trees a given consistent QL
schedule $S$ is defined, and what the properties are of the
document trees that result from a QL schedule. We do this by
defining the sets $N^{min}_I(S)$, $N^{max}_I(S)$, $E^{min}_I(S)$
and $E^{max}_I(S)$, whose informal meaning is respectively the set
of nodes that are required in the input {\sc dt}s on which
$S$ is defined, the set of nodes that are allowed, the set of
edges that are required and the set of edges that are allowed. In
the same way we define the sets $N^{min}_O(S)$, $N^{max}_O(S)$,
$E^{min}_O(S)$ and $E^{max}_O(S)$ taking into account the output
{\sc dt}s.

\begin{definition}
\label{basic-in} Let $S$ be a QL schedule. $\phi_S (n,o)$ ($\phi_S
((m,l,n),o)$) indicates that the first occurrence of the node $n$
(the edge $(m,l,n)$) in the schedule $S$ has the form of the
operator $o$. \footnote{For example, $\phi_S
(n_2,\add(r,l_2,n_2))$ holds in the consistent QL schedule in
Example~\ref{ex:corrnotcorr} above.} $\lambda_S (n,o)$ ($\lambda_S
((m,l,n),o)$) indicates that the last occurrence of the node $n$
(the edge $(m,l,n)$) in the QL schedule $S$ has the form of the
operation $o$. We define the sets $N^{min}_I(S)$, $N^{max}_I(S)$,
$E^{min}_I(S)$ and $E^{max}_I(S)$, and the sets $N^{min}_O(S)$,
$N^{max}_O(S)$, $E^{min}_O(S)$ and $E^{max}_O(S)$ as in
Figure~\ref{basicinoutsets}. \\
A {\sc dt} $T$ is called a basic input tree (basic output tree) of
$S$ iff it contains all the nodes of $N^{min}_I(S)$
($N^{min}_O(S)$), only nodes of $N^{max}_I(S)$ ($N^{max}_O(S)$),
all the edges of $E^{min}_I(S)$ ($E^{min}_O(S)$) and only edges of
$E^{max}_I(S)$ ($E^{max}_O(S)$).
\end{definition}
\begin{figure}[ht]
\framebox{
\parbox[l]{20.0cm}{
\begin{tabbing}
$N^{min}_I(S)$ \= $ = $ \= $\{m \ \vert\  \phi_S (m,\add(m,l,n))\}
\cup
  \{ m\ \vert\ \phi_S (m,\del(m,l,n))\}\ \cup
  \{ n\ \vert\ \phi_S (n,\del(m,l,n))\}$ \\
$N^{max}_I(S)$ \> $ = $ \> $\mathcal{N} - \{n \ \vert\  \phi_S (n,\add(m,l,n))\}$ \\
$E^{min}_I(S)$ \> $ = $ \> $\{(m,l,n) \ \vert\ $ \= $\phi_S
((m,l,n),\del(m,l,n))\}$ \\
$E^{max}_I(S)$ \> $ = $ \> $E^{min}_I(S)\ \cup\ \{(m,l,n)\ \vert\
\mathrm{\ no\ } (m_1,l_1,m) \mathrm{\ nor\ }
(m_1,l_1,n) \mathrm{\ occurs\ in\ } S \}$ \\
$N^{min}_O(S)$ \> $=$ \> $\{m \ \vert\  \lambda_S (m,\del(m,l,n))\}
\cup \{ m\ \vert\ \lambda_S (m,\add(m,l,n))\} \cup\{ n\ \vert\ \lambda_S
(n,
\add(m,l,n))\}$ \\
$N^{max}_O(S)$ \> $=$ \> $\mathcal{N} - \{n \ \vert\  \lambda_S (n,\del(m,l,n))\}$ \\
$E^{min}_O(S)$ \> $=$ \> $\{(m,l,n) \ \vert\ $ \= $\lambda_S((m,l,n),\add(m,l,n))\}$ \\
$E^{max}_O(S)$ \> $ = $ \> $E^{min}_O(S) \cup\ \{(m,l,n)\ \vert\
\mathrm{\ no\ } (m_1,l_1,m) \mathrm{\ nor\ } (m_1,l_1,n) \mathrm{\
occurs\ in\ } S \}$
\end{tabbing}
}}
\caption{The Definition of the basic input and output sets.}
\label{basicinoutsets}
\end{figure}
Consider 
$S = (\add(n_1,l_1,n_2),t_1)$, $(\del(n_4,l_2,n_3),t_2)$,
$(\del(n_1,l_1,n_4),t_3)$ then
\begin{tabbing}
$N^{min}_I(S)\ $\= $=\{n_1,n_3,n_4\}$ \\
$N^{max}_I(S)$\> $=\mathcal{N}-\{n_2\}$ \\
$E^{min}_I(S)$\> $=\{(n_4,l_2,n_3),(n_1,l_1,n_4)\}$ \\
$E^{max}_I(S)$\> $=E^{min}_I(S) \ \cup
 \{ (m,l,n) \in$ \\ \> $\ \ \ \ \ \mathcal{N}\times\mathcal{L}\times\mathcal{N}\ |\  m,n \not= n_2,n_3,n_4\}$ \\
$N^{min}_O(S)$ \> $=\{n_1,n_2\}$ \\
$N^{max}_O(S)$ \> $=\mathcal{N}-\{n_3,n_4\}$ \\
$E^{min}_O(S)$ \> $=\{(n_1,l_1,n_2)\}$ \\
$E^{max}_O(S)$ \> $=E^{min}_O(S)\ \cup \{ (m,l,n) \in$ \\ \> $\ \ \ \ \ \mathcal{N}\times\mathcal{L}\times\mathcal{N}\ |\  m,n \not= n_2,n_3,n_4\}$
\end{tabbing}

We will prove in Theorem~\ref{consistent} that  the application of a
consistent schedule $S$ is defined on each basic input tree of $S$. 

Although $N^{max}_I(S)$, $E^{max}_I(S)$, $N^{max}_O(S)$ and
$E^{max}_O(S)$ are in general infinite, they can be represented in
a finite way: $N^{max}_I(S)$ by $\{n \ |\  \phi_S
(n,\add(m,l,n))\}$, $E^{max}_I(S)$ by $E^{min}_I(S) \cup \{n\ |\
\mathrm{\ there\ is\ a\ } (m_1,l_1,n) \mathrm{\ that\ occurs\ in\
} S \}$, $N^{max}_O(S)$ by $\{n \ |\  \lambda_S
(n,\del(m,l,n))\}$, $E^{max}_O(S)$ by $E^{min}_O(S) \cup \{n\ |\
\mathrm{\ there\ is\ a\ } (m_1,l_1,n) \mathrm{\ that\ occurs\ in\
} S \}$.

\begin{lemma}\label{comp1}
Let $S$ be a schedule with size $n_S$. 
$N^{min}_I(S), N^{max}_I(S), E^{min}_I(S),E^{max}_I(S),$
$N^{min}_O(S), N^{max}_O(S), E^{min}_O(S)$ and
$E^{max}_O(S)$ can be calculated in $O(n_S.log(n_S))$-time and in
$O(n_S)$-space. For each of these sets and for any node or edge it
is decidable in $O(n_S)$-time and $O(log(n_S))$-space whether the
node or edge is in the set.
\end{lemma}
\begin{proof} (Sketch) We can decide whether a node or an edge is
in one of the basic input or output sets by examining the
actions of the schedule $S$.
\end{proof}

When a QL schedule is inconsistent this is always because two
operations in the QL schedule interfere, as for example the  two
operations in the inconsistent transaction of
Example~\ref{ex:corrnotcorr}: $(\add(n_1,l_1,n),t_1)$ and
$(\add(n_2,l_2,n),t_1)$. If these two operations immediately
follow each other then at least one of them will always fail.
However, if between them we find the action $\del(n_1,l_1,n)$ then
this does no longer hold. The following definition attempts to
identify such pairs of interfering operations and states which
operations we should find between them to remove the interference.

\begin{definition} A QL schedule fulfills the {\em C-condition}
iff 
\begin{enumerate}
\item If $\add(n,l_1,n_1)$ and $\add(n_2,l_2,n)$
appear in that order in $S$ then $\del(n,l_1,n_1)$
appears between them. 
\item If $\add(n_1,l_1,n)$ and $\add(n_2,l_2,n)$
appear in that order in $S$ then $\del(n_1,l_1,n)$
appears between them. 
\item If $\add(n,l_1,n_1)$ and $\del(n_2,l_2,n)$
appear in that order in $S$ then $\del(n,l_1,n_1)$
appears between them. 
\item If $\add(n_1,l_1,n)$ and $\del(n,l_2,n_2)$
appear in that order in  $S$ then $\add(n,l_2,n_2)$
appears between them.
\item If $\add(n_1,l_1,n)$ and $\del(n_2,l_2,n)$
appear in that order in  $S$ and $(n_1,l_1)\not=(n_2,l_2)$ then
$\del(n_1,l_1,n)$ appears between them.
\item  If $\del(n,l_1,n_1)$ and $\add(n_2,l_2,n)$
appear in that order in  $S$ then some $\del(n_3,l_3,n)$
appears between them.
\item  If $\del(n_1,l_1,n)$ and $\add(n,l_2,n_2)$
appear in that order in  $S$ then some $\add(n_3,l_3,n)$
appears between them.
\item If $\del(n_1,l_1,n)$ and $\del(n,l_2,n_2)$
appear in that order in  $S$ then some $\add(n_3,l_3,n)$
appears between them.
\item If $\del(n_1,l_1,n)$ and $\del(n_2,l_2,n)$
appear in that order in $S$ then $\add(n_2,l_2,n)$
appears between them.
\end{enumerate}
\end{definition}
The following theorem establishes the relationship between
consistency, basic input trees and the C-condition.

\begin{theorem}\label{consistent}
The following conditions are equivalent for a QL schedule $S$:
\begin{enumerate}
\item there is a basic input tree of $S$ and the application of
$S$ is defined on each basic input tree of $S$. 
\item there is a basic input tree of
$S$ on which the application of $S$ is defined; 
\item $S$ is consistent; 
\item $S$ fulfills the C-condition; 
\item there is a tree on which the application of $S$ is defined
and all trees on which the application of $S$ is defined are
basic input trees of $S$.
\end{enumerate}
\end{theorem}

\begin{proof} (Sketch) Clearly $1 \to 2 \to 3 \to 4$ and $5 \to 3$.
We prove that 4 implies 1. First we prove that there is a
basic input tree for which $S$ is defined. Then we prove that
the application of $S$ is defined on each basic input tree of $S$
by induction on the length of $S$.
Finally 3 implies 5. Indeed, let $S$ be defined on $T$, where $T$
is not a basic input tree of $S$. $T$ does not satisfy one of the
four conditions of Definition~\ref{basic-in}. In each case this
yields a contradiction. 
\end{proof}

\begin{corollary}\label{dec.corr}
It is decidable whether a QL schedule or a transaction is consistent
in $O(n^3_S)$-time and $O(n_S)$-space.
\end{corollary}

\begin{proof} (Sketch) This follows from the decidability of the
C-condition and Theorem~\ref{consistent}.
\end{proof}
For the basic input and output sets we can derive the
following property:

\begin{property}\label{prop1}  If $S$ is a consistent QL schedule then
$E^{min}_I(S)$ and $E^{min}_O(S)$ are forests.
\end{property}
By $\ADD(S)$ we denote the set of edges that are added by the QL
schedule $S$, i.e., they are added without being removed again
afterwards, and by $\DEL(S)$ we denote the set of edges that are
deleted by the QL schedule $S$, i.e., they are deleted without
being added again afterwards.

\begin{definition} \label{adddel}
  Let $S$ be a consistent QL schedule. We denote
  \begin{small}
  \begin{tabbing}
  $\ADD(S) = \{(m,l,n)\ |\  \lambda_S ((m,l,n),\add(m,l,n))\}$ \\
  $\DEL(S) = \{(m,l,n)\ |\  \lambda_S ((m,l,n), \del(m,l,n))\}$
  \end{tabbing}
  \end{small}
We call $\ADD(S)$ the addition set of $S$ and $\DEL(S)$ its
deletion set.
\end{definition}
Remark that two consistent QL schedules with the same $\ADD$ and
$\DEL$ are not necessarily equivalent. Indeed $S_1=
(\del(n_1,l_1,n_2),t_2)$ and $S_2=$
$(\add(n_1,l_1,n_2),t_1),$ $(\del(n_1,l_1,n_2),t_2)$ are not equivalent
although $\ADD(S_1)=$ $\ADD(S_2)$ and $\DEL(S_1)=$ $\DEL(S_2)$.

\begin{lemma}\label{SAD}
  Let $S$ be a consistent QL schedule and $T$ be a basic input tree of $S$.
  $S[T]= T \cup \ADD(S) - \DEL(S)$ is a
  basic output tree\footnote{We consider a graph as the set of its edges and vice versa.}.
\end{lemma}

\begin{proof} (Sketch) Clearly $T \cup \ADD(S) - \DEL(S)$ is the
result of the application of $S$ on $T$. We verify that $T \cup
\ADD(S) - \DEL(S)$ is a basic output tree. 
\end{proof}
The following lemma establishes the relationships between
the addition and deletion sets, and the basic input and output
sets.

\begin{lemma}\label{addel}
Let $S$ be a consistent QL schedule.
\begin{tabbing}
$N^{min}_O = $\= (\= $N^{min}_I - \{n\ |\ \exists (m,l,n)\in \DEL(S)\})\ \cup$ \\
\> $\{n\ |\ \exists (m,l,n)\in \ADD(S)\}$ \\
$N^{max}_O = $\= (\= $N^{max}_I - \{n\ |\ \exists (m,l,n)\in \DEL(S)\})\ \cup$ \\
\> $\{n\ |\ \exists (m,l,n)\in \ADD(S)\}$ \\
$E^{min}_O = (E^{min}_I -  \DEL(S))\cup \ADD(S)$ \\
$E^{max}_O = (E^{max}_I -  \DEL(S))\cup \ADD(S)$
\end{tabbing}
\end{lemma}
\begin{proof} (Sketch) Results from Lemma~\ref{SAD}. 
\end{proof}

\section{Equivalence and Serializability of QL Schedules}

The purpose of a scheduler is to interleave requests by processes
such that the resulting schedule is serializable. This can be done
by deciding for each request whether the schedule extended with the
requested operation is still serializable, without looking at the
instance. In this section we discuss the problem of deciding
whether two consistent QL schedules are equivalent, and
whether a consistent QL schedule is serializable.

To begin with, it can  be shown that the application two QL schedules
over the same set of transactions on the same {\sc dt} $T$ result in
the same {\sc dt}, if they are both defined.
\begin{lemma}\label{eq0}
  Let $S$ and $S'$ be two QL schedules over the same set of
  transactions. $S[T]=S'[T]$ if $S[T]$ and $S'[T]$ are both defined.
\end{lemma}
\begin{proof} (Sketch) Considering a given edge, this edge is alternatively
added and deleted in each of the applications. Since the two
QL schedules are over the same set of transactions, the edge belongs
to no result or to both results 
\end{proof}
As a consequence the problem of deciding whether two consistent
schedules over two given transactions are equivalent reduces to the
problem of deciding whether their result is defined for the same
{\sc dt}s, which can be decided with the help of the basic input
and output sets.

\begin{theorem}\label{eq1}
  Two consistent QL schedules $S_1$, $S_2$ over the same set of transactions are
  equivalent iff they have the same set of basic input trees, ie.
 iff $N^{min}_I(S_1)=N^{min}_I(S_2)$,
  $N^{max}_I(S_1)=N^{max}_I(S_2)$, $E^{min}_I(S_1)=E^{min}_I(S_2)$ and
  $E^{max}_I(S_1)=E^{max}_I(S_2)$.
  Hence their equivalence is decidable in $O(n_S.log(n_S))$-time and
  $O(n_S)$-space.
\end{theorem}

\begin{proof} From Lemma~\ref{comp1} and Lemma~\ref{eq0}. 
\end{proof}
Note that this theorem does not hold for two arbitrary QL
schedules. Indeed $S_1= (\add(m,l,n),t)$ and $S_2=
(\add(m,l,n),t),(\del(m,l,n),t)$ have the same basic input trees
and are not equivalent.

We can use the basic input and output sets to decide whether
one consistent schedule can directly follow another consistent schedule
without resulting in an inconsistent schedule.

\begin{lemma}\label{succ} Let $S_1$ and $S_2$ be two consistent QL
schedules. Let $n_S$ be the size of $S_1.S_2$.
$S_1.S_2$ is consistent iff $N^{min}_I(S_2)\subseteq N^{max}_O(S_1)$,
$E^{min}_I(S_2)\subseteq E^{max}_O(S_1)$, $N^{min}_O(S_1)\subseteq
N^{max}_I(S_2)$, $E^{min}_O(S_1)\subseteq E^{max}_I(S_2)$. The
consistency of $S_1.S_2$  is decidable in $O(n_S.log(n_S))$-time
and $O(n_S)$-space.
\end{lemma}
\begin{proof} (Sketch) A result of the C-conditions, Lemma~\ref{comp1} and Theorem~\ref{consistent}.

\end{proof}
The following lemma shows how the basic input and output sets
can be computed for a concatenation of schedules if we know these
sets for the concatenated schedules.

\begin{lemma}\label{ser}
Let $S_1,S_2,...,S_n$ and $S_1.S_2...S_n$ be $(n+1)$ consistent
QL schedules. Then

$N^{min}_I(S_1...S_n)= {\bigcup _{i=1}^n (N^{min}_I(S_i) \cap {\bigcap
_{k<i} N^{max}_I(S_k)})}$

$N^{max}_I(S_1...S_n)= {\bigcap _{i=1}^n (N^{max}_I(S_i) \cup {\bigcup
_{k<i} N^{min}_I(S_k)})}$

$E^{min}_I(S_1...S_n)= {\bigcup _{i=1}^n (E^{min}_I(S_i) \cap {\bigcap
_{k<i} E^{max}_I(S_k)})}$

$E^{max}_I(S_1...S_n)= {\bigcap _{i=1}^n (E^{max}_I(S_i) \cup {\bigcup
_{k<i} E^{min}_I(S_k)})}$

If $n_S$ is the size of $S_1.S_2...S_n$ then these
equalities can be verified in $O(n_S^3)$-time and $O(n_S)$-space.
\end{lemma}
\begin{proof} By induction using Definition~\ref{basic-in}. 
\end{proof}
Finally, the previous theorems can be used to show that
serializability is decidable.

\begin{theorem}\label{dec.ser}
  Given a QL schedule $S$ of $n_t$ transactions.
  It is decidable whether $S$ is serializable in
  $O(n_t^{n_t}.n_S^3)$-time, and in $O(n_S^2)$-space.
\end{theorem}
\begin{proof} (Sketch)  Indeed,
\begin{enumerate}
    \item we verify whether each transaction is consistent, which is
     done in $O(n_S^3.n_t)$-time and in $O(n_S)$-space
(Corollary~\ref{dec.corr});
    \item we draw a graph that indicates which
transactions can follow directly which other transactions (i.e.
$T_i.T_j$ is defined), which is done in $O(n_t^2.n_S.log(n_S))$-time and
in $O(n_t^2+n_S)$-space (Lemma~\ref{succ});
    \item $S$ is serializable iff there is a Hamilton
path that is equivalent with $S$; to verify this:
    \begin{enumerate}
        \item we calculate the ordered $N^{min}_I$,
$N^{max}_I$,$E^{min}_I$ and $E^{max}_I$ of the transactions, which
is done in $O(n_t.n_S.log(n_S))$-time and $O(n_t.n_S)$-space
(Lemma~\ref{comp1});
        \item there are $O(n_t^{n_t})$ Hamilton paths, for each of them:
        \begin{enumerate}
            \item we  verify its consistency, which is done
 in $O(n_S^3)$-time and $O(n_S)$-space (Corollary~\ref{dec.corr});
            \item we calculate the ordered  $N^{min}_I$, $N^{max}_I$, $E^{min}_I$ and
$E^{max}_I$ of  the Hamilton path, which is done in
$O(n_S.log(n_S))$-time and $O(n_S)$-space (Lemma~\ref{comp1});
            \item Lemma~\ref{ser} and Theorem~\ref{eq1} are verified in $O(n_S^3)$-time and in $O(n_S)$-space.
        \end{enumerate}
     \end{enumerate}
\end{enumerate}

\end{proof}

\section{Equivalence and Serializability of Schedules}
In the previous section we only considered QL schedules,
but in this section we consider all schedules. We start with
generalizing the notions that were introduced for QL schedules.

\begin{definition}
  A schedule $S$ is called
  consistent iff its corresponding QL schedule $S'$ is consistent.
  $\ADD(S)=\ADD(S')$ where $S'$ is the QL schedule of $S$.
  Analogously for $\DEL$, $E^{min}_I$, $E^{max}_I$, $E^{min}_O$, $E^{max}_O$,
  $N^{min}_I$, $N^{max}_I$, $N^{min}_O$, $N^{max}_O$.
\end{definition}

To verify whether two consistent schedules over the same set of
transactions are equivalent, we first eliminate the queries and
verify whether the resulting QL schedules are equivalent. (Cfr.
Theorem~\ref{eq1}). In this section we investigate the equivalence
of two consistent schedules over the same set of transactions and
whose QL schedules are equivalent. In the following examples it is
shown that such schedules can be equivalent on all the {\sc DT}s
they are defined on, on only some of them or on none.

\begin{example}\label{ex.eq.1}
Let $l_1\not= l_3$. Consider the following schedules:
\begin{tabbing}
  $S_1=$ \= $(\add(n_2,l_2,n_3),t_1)$, $(\query(n_1,l_1/l_2),t_2)$, \\
         \> $(\del(n_2,l_2,n_3),t_1)$, $(\del(n_1,l_3,n_2),t_1)$ \\
  $S_2=$ \> $(\query(n_1,l_1/l_2),t_2)$, $(\add(n_2,l_2,n_3),t_1)$, \\
         \> $(\del(n_2,l_2,n_3),t_1)$, $(\del(n_1,l_3,n_2),t_1)$
\end{tabbing}

$S_1$ and $S_2$ are correct and their corresponding QL schedules are
equal. They are equivalent on all {\sc dt}s on which they are defined,
hence they are equivalent. 

Consider the following schedules $S_3$ and $S_4$:
\begin{tabbing}
  $S_3=$ \= $(\add(n_2,l_2,n_3),t_1)$, $(\query(n_1,l_1/l_2),t_2)$, \\
         \> $(\del(n_2,l_2,n_3),t_1)$ \\
  $S_4=$ \> $(\query(n_1,l_1/l_2),t_2)$, $(\add(n_2,l_2,n_3),t_1)$, \\
         \> $(\del(n_2,l_2,n_3),t_1)$
\end{tabbing}
$S_3$ and $S_4$ are consistent and their corresponding QL schedules
are equal. They are equivalent on some {\sc dt}s on which they
are defined and not equivalent on others.  

Finally, let $S_5$ and $S_6$ be the following schedules:
\begin{tabbing}
  $S_5=$ \= $(\add(n_2,l_2,n_3),t_1)$, $(\query(n_1,l_1/l_2),t_2)$, \\
         \> $(\del(n_2,l_2,n_3),t_1)$, $(\del(n_1,l_1,n_2),t_1)$ \\
  $S_6=$ \> $(\query(n_1,l_1/l_2),t_2)$, $(\add(n_2,l_2,n_3),t_1)$, \\
         \> $(\del(n_2,l_2,n_3),t_1)$, $(\del(n_1,l_1,n_2),t_1)$
\end{tabbing}
$S_5$ and $S_6$ are consistent and their corresponding QL schedules
are equal. They are, however, equivalent on no {\sc dt} on which they
are defined.
\end{example}

In order to prove the decidability of the equivalence of two
schedules over the same set of transactions we first define the
notion of $\SOP$, Set Of Prefixes in Subsection 5.1, and some
additional notation in Subsection 5.2.

\subsection{$\SOP$ - Set Of Prefixes}
Informally, the notion ``Set Of Prefixes'' ($\SOP$) of a path
expression $pe$ for a label path $lp$, will allow us to find
a set of path expressions $pe'$, such that all path expressions
$pe'/lp$ together represent exactly these label paths of $pe$
that end on $lp$.
For example, consider the path expression $pe = b//*$ and the
label path $lp = a$. Then $b/a$ and $b//*/a$ represent the
label paths of $pe$ that end with label path $a$. Hence $b$ and
$b//*$ are $a$-prefixes of $b//*$.

We will now define the set of non-empty $lp$-prefixes in $pe$,
denoted as $\SOP(pe)_{lp}$ as a set of path expressions that together
represent  the set of label paths $pe'$  such that $pe'/lp \in
{\rm \bf L}(pe)$\footnote{We consider $pe/\epsilon$ to be equal to
$pe$.}. For instance $\SOP(b//*)_{a} = \{b,b//*\}$.

\begin{definition} \label{SOP0} Let $pe$  be a path expression,
$lp$ be a label path and $l \in \mathcal{L}$. The
 set of non-empty $lp$-prefixes in $pe$, denoted as
$\SOP(pe)_{lp}$  is defined by\\
\\
\begin{tabular}{ll}
$\SOP(pe)_{\epsilon}$ & $= \{pe\}$ \\
$\SOP(pe/*)_{l}$ & $= \SOP(pe/l)_{l}$ = $\{pe\}$ \\
$\SOP(pe//*)_{l}$ & $= \SOP(pe//l)_{l}=\{pe,pe//*\}$ \\
$\SOP(pe/*)_{lp/l}$ & $= \SOP(pe/l)_{lp/l}=\SOP(pe)_{lp}$ \\
$\SOP(pe//*)_{lp/l}$ & $= \SOP(pe//l)_{lp/l}$ \\
& = $\SOP(pe)_{lp}\cup \SOP(pe//*)_{lp}$ \\
Otherwise $\SOP(pe)_{lp}$ & $= \emptyset$. \\
\end{tabular}

Furthermore we define \verb| |${\rm \bf L} (\SOP(pe)_{lp})=\bigcup_{pe_i \in \SOP(pe)_{lp}}{\rm \bf L}(pe_i)$.
\end{definition}

\begin{lemma}\label{SOP9}
${\rm \bf L} (\SOP(pe)_{lp})=\{ lp'\ |\ lp'/lp \in {\rm \bf
L}(pe)\}$. \end{lemma}

\begin{example}\ \\ \vspace{-0.35cm}
\begin{itemize}
  \item $\SOP(a/*/*/b)_{a/b}=\SOP(a/*/*)_{a}= \{a/*\}$
  \item $\SOP(a//*/c)_{a/b/c}=\SOP(a//*)_{a/b}=\SOP(a)_{a} \cup
    \SOP(a//*)_{a}= \{a,a//*\}$
  \item  $\SOP(*//*)_{a/b/c}=\SOP(*)_{a/b}\cup\SOP(*//*)_{a/b}=
  \emptyset \cup \SOP(*)_{a}\cup\SOP(*//*)_{a}=\emptyset \cup
    \emptyset \cup \{*,*//*\}=\{*,*//*\}$
  \item $\SOP(a//b//d)_{b/c/d}= \{a,a//*, a//b, a//b//*\}$
\end{itemize}
\end{example}

\begin{lemma}\label{SOP1} Let $pe$  be a path expression and $lp$ be a
label path. $\SOP(pe)_{lp} = \{ pe'\ |\ pe'$ a prefix of $pe$ and
${\rm \bf L}(pe'/lp) \subseteq {\rm \bf L}(pe)\} \cup \{ pe'//*\
|\ pe'$ a prefix of $pe$ and ${\rm \bf L}(pe'//*/lp) \subseteq
{\rm \bf L}(pe)\}$. 
\end{lemma}

\begin{lemma}\label{SOP2} Let $pe$  be a path expression of length $n_{pe}$ and $lp$ be a
label path. $\SOP(pe)_{lp}$ is uniquely defined, finite and is
computable in $O(n_{pe}^2.(n_{pe} + n_{lp}))$-time and $O(log(n_{pe}+n_{lp}))$-space.
\end{lemma}

\begin{proof}
From Lemma~\ref{SOP1} we know that we have to calculate the two
sets: $\{ pe'\ |\ pe'$ a prefix of $pe$ and ${\rm \bf L}(pe'/lp)
\subseteq {\rm \bf L}(pe)\}$ and $\{ pe'//*\ |\ pe'$ a prefix of
$pe$ and ${\rm \bf L}(pe'//*/lp) \subseteq {\rm \bf L}(pe)\}$.
Hence $\SOP(pe)_{lp}$ can be calculated in $O(n_{pe}^4)$-time
~\cite{miklau02containment}, and in $O(n_{pe}^2)$-space. 
\end{proof}

\begin{lemma}\label{SOP3} Let $pe$  be a path expression and $lp_1$ and $lp_2$ be
two label paths.   ${\rm \bf L} (\SOP(pe)_{lp_1})\subseteq {\rm
\bf L} (\SOP(pe)_{lp_2})$ iff $\forall pe_i \in \SOP(pe)_{lp_1}
({\rm \bf L}(pe_i/lp_2) \subseteq {\rm \bf L}(pe))$.
\end{lemma}

\begin{proof} From Definition \ref{SOP0} and Lemma
\ref{SOP9}.
\end{proof}

\begin{theorem}\label{SOP4} Let $pe$  be a path expression and $lp_1$ and $lp_2$ be
two label paths.  It is decidable in $O(n_{pe}^2.(n_{pe}+n_{lp}))$-time  and in
$O(n_{pe} + log(n_{pe}+n_{lp}))$-space whether ${\rm \bf L} (\SOP(pe)_{lp_1})={\rm
\bf L} (\SOP(pe)_{lp_2})$.
\end{theorem}

\begin{proof} From Lemmas \ref{SOP2} and \ref{SOP3}.

\end{proof}

\subsection{$\PQRN$ - Potential Query Result Nodes }\label{addnotsection}
The main concept that is introduced in this subsection is the set of Potential
Query Result Nodes ($\PQRN$) for a query $Q$ in a schedule $S$. This set will
contain all nodes $n$, that are added or deleted in $S$, and for which there
exists a document tree $T$, such that $n$ is in the result of the query $Q$ when
$S$ is applied to $T$.
\footnote{This notion is only defined for a subset of queries, which will be
specified later on.}.
For this puprose, we need to introduce some additional notations to
characterize the trees on which a query $Q$ in a schedule $S$ will be
executed. We will use these notations later on, and we also give some
complexity results for calculating the value of these concepts.

Let $S$ be a consistent schedule that contains the query $Q = \query(n,pe)$.
\begin{itemize}
    \item We denote by $S^Q$ the actions of $S$  that occur before
    $Q$; $S^Q$ is called a subschedule of $S$;
     \item Let $T$ be a basic input tree of $S$. We define $T^Q=S^Q[T]$  as the {\sc dt} on
     which $Q$ in $S$  is evaluated; hence the result of the application of the query $Q$ in $S$ is $Q[T^Q]$;
    \item We denote by $E^{min}(S^Q)$ as the set
that contains exactly those edges that {\bf are required} in
$T^Q$; This set is equal to $(E^{min}_I(S)- \DEL(S^Q)) \cup
\ADD(S^Q)$ (Lemma~\ref{addel});
    \item We denote by $E^{max}(S^Q)$ as the set
that contains exactly those edges that {\bf are allowed} in $T^Q$;
This set is equal to $(E^{max}_I(S) - \DEL(S^Q))\cup \ADD(S^Q)$
(Lemma~\ref{addel}).
\end{itemize}

$E^{min}(S^Q)$ is a forest (Property~\ref{prop1}). As such every
node $m$ of $E^{min}(S^Q)$ has a unique ancestor without a parent
in $E^{min}(S^Q)$; it is denoted by $ARoot(S^Q,m)$. The label of
the path from $ARoot(S^Q,m)$ to $m$ in $E^{min}(S^Q)$ is denoted
by $ALabel(S^Q,m)$.

\begin{lemma}\label{labro}
$Alabel(S^Q,m)$ and $ARoot(S^Q,m)$ can be computed in
$O(n_S^2)$-time and $O(n_S)$-space.
\end{lemma}
\begin{proof} A consequence of Lemma~\ref{comp1}.

\end{proof}
If $\add(m, l, n)$ or $\del(m, l, n)$ are  operations of $S$
we say that $n$ is a {\em non-building-node} of $S$. Otherwise $n$
is called a {\em building-node} of $S$. Note that \\
$E^{max}(S^Q) = E^{min}(S^Q) \cup \{$edges that contain only building nodes$\}$ since \\
$E^{min}(S^Q) =(E^{min}_I(S)- \DEL(S^Q)) \cup \ADD(S^Q)$,\\
$E^{max}(S^Q) = (E^{max}_I(S) - \DEL(S^Q))\cup \ADD(S^Q)$ and \\
$E^{max}_I(S) = E^{min}_I(S)\cup \{$edges that contain only building nodes$\}$.

We will now define the set of nodes $\PQRN(S,Q)$. This set will contain all non-building-nodes that can be in the result of a query that starts with a node $n$ that is not in $E^{min}(S^Q)$. After the formal definition we will show that this definition corresponds to this informal description. Finally we will show that this set is computable in polynomial time and space.

\begin{definition} Let $S$ be a consistent schedule that contains a query $Q=\query(n,pe)$. We define the set $\PQRN(S,Q)$ as:

$\PQRN(S,Q) = \{ m \vert\ $
\begin{itemize}
    \item $m$ a node in the graph $E^{min}(S^Q)$;
    \item $m$ a non-building-node;
    \item $ARoot(S^Q,m)$ a building-node;
    \item $ARoot(S^Q,m) \not= n$;
    \item ${\rm \bf L} (\SOP(pe)_{ALabel(S^Q,m)})\not=\emptyset$
\end{itemize}
$\}$.
\end{definition}

\begin{lemma}\label{PQRNcorrect}
Let $S$ be a consistent schedule, $Q = \query(n,pe)$ a query that appears in $S$, and $n$ a node that is not in the graph $E^{min}(S^Q)$. Then $\PQRN(S,Q)$ is the set of non-building-nodes $m$, such that there exists a basic input tree $T$ of $S$ for which $m$ is in the result of the query $Q$ on the document tree $S^Q[T]$.
\end{lemma}

\begin{lemma}\label{PQRN}
$\PQRN(S,Q)$  can be computed in $O(n_S^5)$-time and $O(n_S)$-space.
\end{lemma}

\begin{proof}
From Theorem~\ref{SOP4} and Lemma~\ref{labro}.
\end{proof}

\subsection{Decidability of Equivalence}
We will now establish the main result of this paper by proving
that the equivalence of two schedules is decidable in our
framework.

\begin{lemma}\label{DE1} Given two consistent schedules
$S_1$ and $S_2$ over the same set of transactions and whose QL
schedules are equivalent. Let $Q=\query(n,pe)$ be a query in these
schedules and let $n_a$ be the total number of actions in $S_1$
and $S_2$. It is decidable in $O(n_S^6)$-time and
$O(n_S)$-space whether $Q$ gives the same answer in $S_1$ as in
$S_2$ for every possible basic input tree of $S_1$ and $S_2$.
\end{lemma}

\begin{proof}
 The next condition $CND(S_1,S_2,Q)$ detects when  $Q$ gives
 the same answer in $S_1$ as in $S_2$ for every
possible basic input tree of $S_1$ and $S_2$ : \\
{\bf Definition of $CND(S_1,S_2,Q)$}
\begin{enumerate}
    \item  $\{ m\ |\ $ there is a path of {\rm \bf L}(pe) from $n$
to $m$ in $E^{min}(S^Q_1) \} = \{ m\ |\ $ there is a path of {\rm
\bf L}(pe) from $n$ to $m$ in $E^{min}(S^Q_2) \}$; this can be
done in $O(n_S^3)$ time; this is a consequence of a result
in \cite{miklau02containment}
    \item furthermore, if {\bf $n$ is a building-node of $S_i$}:
\begin{enumerate} \item
   $PQRN(S_1,Q)=PQRN(S_2,Q)$
\item for the nodes $m \in PQRN(S_1,Q)$  hold that
    \begin{enumerate}
        \item $ARoot(S^Q_1,m) = ARoot(S^Q_2,m)$
        \item ${\rm \bf
L} (\SOP(pe)_{ALabel(S^Q_1,m)})={\rm \bf L}
(\SOP(pe)_{ALabel(S^Q_2,m)})$
     \end{enumerate} \end{enumerate}
All this can be computed in $O(n_S^6)$-time and in $O(n_S)$-space.
\end{enumerate}
\end{proof}

The definition of the $CND$ condition is illustrated in the following
example.

\begin{example} In Example~\ref{ex.eq.1} we have 
\begin{itemize}
\item $E^{min}(S^Q_1)=\{(n_1,l_3,n_2),(n_2,l_2,n_3)\}$ and
$E^{min}(S^Q_2)=\{(n_1,l_3,n_2)\}$; $1.$ is fulfilled; $n_1$ is a
building-node; $n_2$ and $n_3$ are non-building-nodes;
$\PQRN(S_1,Q)= \PQRN(S_2,Q) = \emptyset$; hence $CND(S_1,S_2,Q)$ is fulfilled
and $Q$  gives the same answer in $S_1$ as in $S_2$ for every possible
basic input tree of $S_1$ and $S_2$.

\item $E^{min}(S^Q_3)=\{(n_2,l_2,n_3)\}$ and $E^{min}(S^Q_4)=\emptyset$; $1.$
is fulfilled; $n_2$ is a building-node; $n_3$ is a non-building-node;
$\PQRN(S_3,Q)=\{n_3\}$ and $\PQRN(S_4,Q) = \emptyset$; hence $2.(a)$ is not
fulfilled and $Q$ does not give the same answer in $S_3$ as in $S_4$ for
every possible basic input tree of $S_3$ and $S_4$.

\item $E^{min}(S^Q_5)=\{(n_1,l_1,n_2),(n_2,l_2,n_3)\}$ and 
$E^{min}(S^Q_6)=\{(n_1,l_1,n_2)\}$; hence $S_5$ and $S_6$ are not equivalent,
since $1.$ is not fulfilled and $Q$ does not give the same answer in $S_5$ as
in $S_6$ for every possible basic input tree of $S_5$ and $S_6$.
\end{itemize}
\end{example}

\begin{theorem}\label{DE2} Given two consistent schedules
$S_1$ and $S_2$ over the same set of transactions and whose QL
schedules are equivalent.  It is decidable in
$O(n_S^6)$-time and $O(n_S)$-space whether they are
equivalent.
\end{theorem}
\begin{proof}
Consequence of Lemma~\ref{DE1}. 
\end{proof}

Finally, we can now combine the previous theorems to show that
serializability is decidable in our framework.

\begin{theorem} Given a consistent schedule
$S$.  It is decidable in
$O(n_t^{n_t}.n_S^6)$-time and $O(n_S^2)$-space
whether $S$ is serializable.
\end{theorem}
\begin{proof}
From Theorem~\ref{dec.ser} and Theorem~\ref{DE2}.
\end{proof}

\section{Conclusion and Future Work}
In this paper we have presented a concurrency control mechanism for
semistructured databases. This mechanism is document-independent in the
sense that two schedules of semistructured transactions are equivalent
iff they are equivalent on all possible documents. This notion of
equivalence is a special form of view equivalence. The transactions
that we consider, consist of simple updates (inserting and deleting
edges at the bottom of a tree) and queries (simple path expressions
containing child and descendant steps).
We have shown that equivalence of schedules can be decided
efficiently (i.e., in polynomial time in the size of the schedule),
and that the serializability can be decided in time polynomial in the
size of the schedule and exponential in the number of transactions.
Improving this complexity result is expected to be difficult, since
it is generally known that deciding view serializability is
$NP$-complete \cite{WV02}.

In future work, we will extend the results of this paper by defining
the behaviour of currently undefined actions, and hence allowing
more schedules to be serialized.
For example, the addition of an edge which is already in the input
tree is undefined in our current work, and hence the operation fails.
However, we could also say that as a result of this addition,
we obtain an output tree which is equal to the input tree, and a
message which indicates that the edge was already present.
In this approach the result of a schedule applied on a document
tree would be an annotated version of the schedule and an
output document tree. A schedule would then be serializable iff there
exists a serial schedule with the same operations, which has, for each
document, the same output document tree and the same message for
each operation.

\bibliographystyle{abbrv}
\bibliography{biblock}

\end{document}